\documentclass{article}
\usepackage[a4paper,left=20mm,right=20mm,top=15mm,bottom=15mm]{geometry}
\usepackage{url}
\usepackage{changepage}
\usepackage{booktabs}
\usepackage{siunitx}
\usepackage{rotating}
\usepackage{array,multirow}
\usepackage{pifont}
\newcommand{\cmark}{\ding{51}}
\newcommand{\xmark}{\ding{55}}

\AtBeginDocument{%
  \providecommand\BibTeX{{%
    \normalfont B\kern-0.5em{\scshape i\kern-0.25em b}\kern-0.8em\TeX}}}

\begin{document}

\title{A Critical Review on the Use (and Misuse) of Differential Privacy in Machine Learning}

\author{Alberto Blanco-Justicia$^1$, David S\'anchez$^1$, 
Josep Domingo-Ferrer$^1$,\\
and Krishnamurty Muralidhar$^2$}
\date{$^1$ Universitat Rovira i Virgili, 
Dept. of Computer Science and Mathematics,
        UNESCO Chair in Data Privacy,
        CYBERCAT-Center for Cybersecurity Research of Catalonia,
        Av. Pa\"isos Catalans 26, 43007 Tarragona, Catalonia\\
E-mail \{alberto.blanco,david.sanchez,josep.domingo\}@urv.cat\\
$^2$ University of Oklahoma,
Dept. of Marketing and Supply Chain Management, 307 West Brooks, Adams Hall Room 10, Norman, OK 73019, U.S.A.\\
E-mail krishm@ou.edu}



\maketitle

\begin{abstract}
We review the use of differential privacy (DP) for privacy protection in machine learning (ML). We show that, driven by the aim of preserving the accuracy of the learned models, DP-based ML implementations are so loose that they do not offer the {\em ex ante} privacy guarantees of DP. Instead, what they deliver is basically noise addition similar to the traditional (and often criticized) statistical disclosure control approach. Due to the lack of formal privacy guarantees, the actual level of privacy offered must be experimentally assessed {\em ex post}, which is done very seldom. In this respect, we present empirical results showing that standard anti-overfitting techniques in ML can achieve a better utility/privacy/efficiency trade-off than DP.
\end{abstract}

%



\section{Introduction}
\label{intro}

As long ago as the 1970s, official statisticians~\cite{Dale77} began to worry about potential disclosure of private information on people or companies linked to the publication of statistical outputs. This ushered in the statistical disclosure control (SDC) discipline~\cite{hund12}, whose goal is to provide methods for data anonymization. Also related to SDC is randomized response (RR, \cite{Warn65}), which was designed in the 1960s as a mechanism to eliminate evasive answer bias in surveys and  turned out to be very useful for anonymization. 
The usual approach to anonymization in official statistics is {\em utility-first}: anonymization parameters are iteratively tried until a parameter choice is found that  preserves sufficient analytical utility while reducing below a certain threshold the risk of disclosing confidential information on specific respondents. Both utility and privacy are evaluated {\em ex post} by respectively measuring the information loss and the probability of re-identification of the anonymized outputs. 

In the late 1990s, in the wake of the explosion of the web and the consequent surge of data collection and exchange beyond official statistics, privacy protection became a mainstream topic in the computer science community, which introduced a different angle, namely {\em privacy-first} data protection. 
In this approach, a {\em privacy model} specifying an {\em ex ante} privacy condition is enforced using one or several SDC methods, such as noise addition, generalization or microaggregation~\cite{models}. The parameters of the SDC methods thus depend on the privacy model parameters, and too strict a choice of the latter may result in poor utility. Privacy models have the advantage of setting the privacy level by design, without having to empirically evaluate it.  
The first privacy model was $k$-anonymity~\cite{Sama98}, which was followed by a number of extensions, such as $l$-diversity~\cite{ldiver}, $t$-closeness~\cite{tclose} and others.
All privacy models in the $k$-anonymity family were designed to anonymize microdata sets, that is, data sets formed by records that correspond to individual respondents. 

{\em Differential privacy} (DP, \cite{dwor06}) is a privacy model that arose later, in 2006, in the cryptographic community. Unlike $k$-anonymity, DP was designed to anonymize the outputs of interactive queries to a database. 
Given $\epsilon \geq 0$, a randomized query function $\kappa$ (that returns the exact answer to a query function $f$ plus some noise) satisfies $\epsilon$-DP if, for all data sets $D_1$ and $D_2$ that differ in one record and all $S \in Range(\kappa)$, it holds that
\begin{equation}
\label{dpcondition}
\Pr(\kappa(D_1)\in S) \leq \exp(\epsilon) \times \Pr(\kappa(D_2)\in S).
\end{equation}
In plain words, the presence or absence of any single record in the database must not be noticeable from the query answers, up to a factor exponential in $\epsilon$. 
If each record corresponds to a different individual respondent, this means that the individual's information stays confidential. The smaller $\epsilon$, also known as the {\em privacy budget}, the higher the protection. The amount of noise that $\kappa$ must incorporate to satisfy DP is inversely proportional to $\epsilon$ and directly proportional to the global sensitivity $\Delta(f)$, that is, to how much the output of the query function $f$ can change due to the addition, deletion or change of a single record in the database. Mild noise may suffice when $f$ is a rather insensitive function such as the median or the mean, but much larger noise may be needed for more sensitive query functions such as maxima or minima. 

DP has several advantages w.r.t. privacy models in the $k$-anonymity family and the utility-first approach:
\begin{itemize}
\item DP provides a strong privacy guarantee that
is largely independent of the attacker's background knowledge. In contrast, in $k$-anonymity an assumption on the attacker's knowledge is implicit in the choice of quasi-identifier attributes. However, as noted 
in~\cite{Kife2011} and~\cite{clif2013}, DP is not entirely free of assumptions either. 
\item DP has very nice composability properties. {\em Sequential composition} states that, if the outputs of queries $\kappa_i$, for $i=1,\ldots,m$, on non-independent data sets are individually protected under $\epsilon_i$-DP, then the output obtained by composing all individual query outputs is protected under $\sum_{i=1}^m \epsilon_i$. This formalizes the intuition that, the more outputs are accumulated on overlapping data, the weaker is the extant protection. {\em Parallel composition} states that, in the specific case that $m$ query outputs were computed on $m$ disjoint and independent data sets and protected under $\epsilon$-DP, then the composition of those outputs is still protected under $\epsilon$-DP. 
\end{itemize}

To understand the practical protection offered by a certain privacy budget $\epsilon$,  
let us use the connection between DP and the aforementioned randomized
response. Consider RR for a binary attribute,
so that the reported randomized answer is equal to the true
answer with probability $p\geq 0.5$ and different with probability
$1-p$. 
It is shown in \cite{wang14,wang16} that, if 
$p= \exp(\epsilon)/(1 + \exp(\epsilon))$, then 
binary RR satisfies $\epsilon$-DP. For example, DP with $\epsilon=8$ is satisfied by
RR with $p=0.9996646$, that is, by RR reporting
the true answer with probability practically 1
(which basically amounts to no disclosure protection
being offered).
In this sense, the inventor of DP recommends setting $\epsilon \leq 1$~\cite{dwork11}, so 
that the practical protection obtained comes reasonably close to the theoretical privacy guarantee.
Actually, one of the DP co-inventors is reported to have stated that using
values of $\epsilon$ as high as 14 is pointless in terms of privacy~\cite{greenberg}.
This contrasts with other parameterized privacy models, for which any value of the privacy parameter is privacy-meaningful. For instance, with $k$-anonymity, the re-identification risk is upper-bounded by $1/k$ for any positive integer $k$.

The downside of DP in interactive settings is that it limits the number
of queries that can be usefully answered.
If $m$ queries are to be answered on the same subject's data, sequential composition requires the overall budget $\epsilon$ to be split for the $m$ queries. In case of equal splitting, each query has budget $\epsilon/m$, which entails an $m$-fold noise increase, with the subsequent utility decrease. Moreover, since noise is calibrated to the output sensitivity of a particular query, queries that have higher sensitivity cannot be answered. 
See~\cite{clif2013} for additional criticisms to DP. 

Nonetheless, due to its convenient properties and strong privacy guarantee, DP was rapidly adopted by the research community, up to the point that many researchers now consider DP the gold standard in privacy protection \cite{edmond}.
Not surprisingly, many works have endeavored to apply DP outside the interactive setting it was designed for. Unfortunately, they often neglect the assumptions, limitations and prerequisites of DP.

Such is the case when using DP to anonymize microdata sets. The record corresponding to a respondent can be viewed as the output of an identity query, that is, a query whose function $f$ merely extracts the value of a certain record from the microdata set. The global sensitivity $\Delta(f)$ of such a single-record query is typically enormous, because the attribute values in the record can change within their entire domains. Hence, a correspondingly enormous amount of noise is required for each identity query to enforce DP with a reasonably safe $\epsilon$. As a result, a DP microdata set is obtained whose records differ very significantly from the 
corresponding original records~\cite{VLDB}. In summary, if DP is properly applied, the utility of microdata releases is seriously decreased. Practical applications of DP for microdata have tried to circumvent this problem through sloppy implementations or unreasonably high privacy budgets $\epsilon$~\cite{acm}. 

DP has been also used by big companies in the context of microdata collection (rather than release), by locally anonymizing data under DP at the source respondent's device. For this to be possible with reasonable utility, Apple has used $\epsilon$ values ranging between 6 and 43~\cite{greenberg}, and Google use $\epsilon$ up to 9 in the RAPPOR 
technology~\cite{Erlingsson2014}. 
Moreover, to escape the use of sequential composition in continuous data collection on the same subject, which would exponentially reduce the utility of the data, Apple considered the data on an individual gathered in different days to be
independent~\cite{greenberg}, and Google considered sequential composition just on values that have not changed from the previous collection~\cite{accessnow}. Both approaches are severely flawed, since it is obvious 
that different pieces of data on the {\em same} individual cannot be viewed as independent~\cite{acm}.  

More recently, DP has been extensively applied to enhance privacy in machine learning (ML). The rationale is that ML requires huge amounts of training data that are most often personal data and thus need to be protected. Protecting the individuals' data used to train ML models or protecting the learned models prior to releasing them bear many similarities with protecting continuous data collection or protecting data releases, respectively. Therefore, the challenges are similar, basically the difficulty of reconciling the strong privacy of DP with enough data utility (model accuracy in the ML context). However, recent advances in the application of DP to ML promise DP-protected ML models with minimum accuracy loss and with reasonably safe DP parameters (and thus with ``enough'' privacy) \cite{Abadi2016}. This seems counterintuitive, given the strong constraints imposed by DP and the privacy implications of releasing an ML model trained on personal data. 

\subsection*{Contributions and plan of this survey}

In this survey we critically review the use of DP in ML. Specifically, we focus on the following assertion by the DP inventor~\cite{dwor19} about the state of the art of DP-based ML:
\begin{quotation}
{\em When meaningfully implemented, DP supports
deep data-driven insights with minimal worst-case privacy loss. When not meaningfully
implemented, DP delivers privacy mostly in name.}
\end{quotation}

We will show that DP-based ML implementations are mainly on the non-meaningful side. Not only because of the use of unreasonably unsafe privacy parameters, but also due to the use of relaxations that should not be counted as DP. These flaws are motivated by the attempt to salvage utility, {\em i.e.} model accuracy, which is of utmost importance in ML and is often incompatible with strict DP. 
In fact, in many implementations reported in the ML literature, the use of DP amounts to little more than adding noise without precise {\em ex ante} privacy guarantees. For this reason, the actual level of privacy achieved must be evaluated empirically, as typically done in the utility-first approach common in official statistics. 

Therefore, in addition to surveying the empirical figures reported by works on DP-protected ML, we also report our own original empirical results on the practical utility/privacy/efficiency achieved by common implementations of DP in ML. 
These show that, as expected, privacy risks in ML models come from fitting too much the (private) training data ({\em i.e.}, {\em overfitting}). Further, it can be seen that DP not only fails to mitigate overfitting better than standard methods employed in ML for that purpose, but it adds very significant computation overhead.

Compared to previous related surveys such as 
\cite{Gong,yin2021}, which systematically and theoretically review the current literature, we offer a critical analysis, both theoretical and empirical, of (the flaws in) the application of DP to ML. Compared to more practical surveys, such as~\cite{Jayaraman2019}, we provide a more up-to-date compilation of works on DP-protected ML  
(including non-centralized ML applications of DP such as federated learning), and we also contribute original experimental work reporting practical accuracy, privacy and computational cost.

The remainder of the paper is organized as follows. Section~\ref{fitting} reviews state-of-the-art strategies used to fit DP for ML. Sections~\ref{centralized} and~\ref{decentralized} review works applying DP to ML, both in centralized and decentralized settings. Section~\ref{empirical} reports our empirical work.
Finally, conclusions are gathered in Section~\ref{conclusions}.

\section{Fitting (or bending) DP for ML}
\label{fitting}

As introduced above, protecting the individuals' data used to train ML models via DP faces similar difficulties as protecting continuous data collection: since successive model training epochs are computed on the same (or at least, not completely disjoint) data, sequential composition applies. For this reason, the effective $\epsilon$ grows with the number of epochs,
which, worse yet, is not known in advance.
This means that a great amount of noise needs to be added at each epoch to keep the effective $\epsilon$ small, but this comes at the expense of the model's accuracy. Moreover, due to the distortion caused by the added noise, the convergence of the model becomes slower, and thus more training epochs are required and, in turn, more noise per epoch must be added to meet a certain target $\epsilon$. A vicious cycle ensues from which it is difficult to escape.     

The amount of noise added (and hence the utility loss) is also directly proportional to the query sensitivity. Thus, a way to avoid increasing $\epsilon$ is to try to reduce the sensitivity of queries, even if this may sound a bit contrived. In ML, two sensitivity-reduction strategies are favored:
\begin{itemize}
    \item {\em Output truncation.} It is a standard practice to truncate gradients in order to decrease their sensitivity~\cite{Abadi2016}, although this may slow down or bias the learning process.
    \item {\em Prior subsampling.} The query output ({\em e.g.} the gradient or model update) is computed on a random sample of the private data set rather than on its entirety. The fact that records with extreme values may not be part of the sample reduces the ``expected'' query sensitivity. However, subsampling may slow down the learning process or bias it. Additionally, Dwork indicates in~\cite{dwork11} that subsampling is insufficient to protect privacy: if an extreme individual happens to be part of the subsample, she or he may be spotted. 
\end{itemize}

Due to the impossibility of obtaining usable models with safe/small-enough $\epsilon$, strict DP has been rarely applied in ML~\cite{Triastcyn2019,Jayaraman2019}. Instead, relaxations of the DP definition are used across the board. By far, the most common relaxation
is $(\epsilon,\delta)$-DP~\cite{dworkourdata06}, which is satisfied by a randomized query function $\kappa$ if, for all data sets $D_1$ and $D_2$ that differ in one record and all $S \in Range(\kappa)$, it holds that
\begin{equation}
\label{deltadpcondition}
\Pr(\kappa(D_1)\in S) \leq \exp(\epsilon) \times \Pr(\kappa(D_2)\in S) + \delta.
\end{equation}
$\delta>0$ is usually called the failure probability, that is, the probability that strict $\epsilon$-DP is not satisfied. However, a close look at Expression (\ref{deltadpcondition}) shows that it is possible to satisfy it always while violating Expression (\ref{dpcondition}). In fact, $(\epsilon,\delta)$-DP amounts to $\epsilon'$-DP, 
for some $\epsilon' > \epsilon$ that depends on $\epsilon$ and $\delta$ (note that $\epsilon'$ must satisfy $\exp(\epsilon')\times \Pr(\kappa(D_2) \in S) =
\exp(\epsilon)\times \Pr(\kappa(D_2) \in S) + \delta$). In~\cite{dwork_privacybook} it is mentioned that one should take $\delta \ll 1/n$, where $n$ is the number of records in the database; otherwise, satisfying $(\epsilon,\delta)$-DP might be compatible with leaking a certain number of records. A usual mechanism to satisfy $(\epsilon,\delta)$-DP is Gaussian noise addition~\cite{dwork_privacybook}. 

Other relaxations of DP are concentrated DP~\cite{ref18_1902.08874}, zero-concentrated 
DP\\ \cite{ref9_1902.08874} and R\'enyi DP~\cite{ref49_1902.08874}. These variants look at the privacy loss as a random variable and bound the tails of its distribution. In so doing, they consider the average rather than the worst-case privacy loss. The result is a refinement of $(\epsilon,\delta)$-DP which turns $\delta$ into a real failure probability. These variants are not broadly used because, among other reasons, their parameters are hard to interpret~\cite{Triastcyn2019}. 
In~\cite{Jayaraman2019} a table is given that expresses these three variants in terms of $(\epsilon,\delta)$-DP. For a more intuitive discussion, see also~\cite{Canonne21}.

The common purpose of $(\epsilon,\delta)$-DP and the other aforementioned variants is to relax $\epsilon$-DP and thereby require less noise addition, which in turn should allow more utility preservation. Yet, the price paid is the possibility of more privacy leakage, as demonstrated in~\cite{Jayaraman2019}.

For a long time, even relaxed DP remained unachievable in many practical ML settings. Some attempts led to unreasonably high $\epsilon$ values~\cite{Shokri2015}, which were shown to be ineffective against attacks~\cite{hitaj17}.
A major breakthrough in the reduction of the required $\epsilon$ was done in~\cite{Abadi2016} with the introduction of the moments accountant method for bounding the cumulative privacy loss across the successive epochs. This method is currently the standard for keeping track of the privacy loss during training due to its practical implementation in the TensorFlow-Privacy library. In comparison with sequential composition, the moments accountant method considers that only a fraction of the individual's training data is employed at each epoch, which allows tighter privacy bounds. With this method, single-digit $\epsilon$ values (barely below 10) can be reached, which are viewed as very good in ML, 
that is, preserving some meaningful utility.

\section{Applying DP to centralized ML}
\label{centralized}

In centralized ML the learning process is entirely managed by a single entity. This manager may apply privacy protection to the {\em input} of the learning process (by acting on the training data or on the objective function), to the {\em intermediate results} of the learning process (by protecting successive model updates), or to the {\em output} of learning (by protecting the learned model prior to releasing it). 

When applying DP to centralized ML, an important distinction is whether the objective function is convex or not. In the convex case, DP can be applied by adding noise to the objective function. The amount of noise needed to satisfy DP requires computing the sensitivity of the ML algorithm~\cite{ref12_1902.08874}.

The non-convex case basically amounts to deep learning with neural networks, which is the most popular approach to ML nowadays. Due to the hidden layers of neural networks, there may be multiple local minima. As noted in~\cite{Jayaraman2019}, the sensitivity analysis methods of~\cite{ref12_1902.08874} cannot be used in this case. As a consequence, adding DP perturbation to the objective function is not an option in deep learning.   
The most popular strategy to achieve DP in deep learning is to add noise to the gradients. The problem is that in deep learning gradients are unbounded, which entails unbounded sensitivity and unbounded noise addition to attain DP. We explained in Section~\ref{fitting} that the usual fix is to manually clip the gradients at each iteration~\cite{Abadi2016}, which bounds sensitivity and also bounds the noise that needs to be added. 

As discussed earlier, the standard approach to DP-based ML consists in enforcing $(\epsilon,\delta)$-DP by clipping updates to a fixed threshold $C$ and then adding Gaussian noise with variance $C^2\sigma^2$. The $\sigma$ parameter is calibrated to bound the privacy loss in each epoch, and the privacy loss ($\epsilon$) is accumulated via the moments accountant method~\cite{Abadi2016}. As introduced in Section~\ref{fitting}, setting the $\epsilon$ parameter in this way is difficult and non-intuitive because it requires predicting the number of epochs beforehand, which not only depends on the training data, but also on the distortion added by the DP mechanism that, in turn, depends on the privacy parameter. 
The $\delta$ parameter is set to be smaller than $1/n$, with $n$ being the number of records in the data set.

\begin{table}
\tiny
  \caption{DP for centralized deep learning: a comparison of recent works}
  \label{tab:ML}
\begin{adjustwidth}{-1.7cm}{}
  \begin{tabular}{llllrrrrr}
    \toprule
    Reference (\textit{cites})&Data set&Size&Original acc.&DP parameters&DP accuracy&Relative accuracy loss \\
    \midrule

    Abadi \emph{et al.} 2016 \cite{Abadi2016} (\textit{2,924}) & CIFAR10 & 50,000 & 86\% & $\epsilon=\{2,4,8\};\delta=10^{-5}$&\{67\%,70\%,73\%\}&\{22.1\%,18.6\%,15.1\%\}\\
    
     Abadi \emph{et al.} 2016 \cite{Abadi2016} (\textit{2,924}) & MNIST & 60,000 & 98.3\% & $\epsilon=\{0.5,2,8\};\delta=10^{-5}$&\{90\%,95\%,97\%\}&\{8.4\%,3.3\%,1.3\%\}\\
     
    Papernot \emph{et al.} 2017 \cite{Pape17} (\textit{657}) & MNIST & 60,000 & 99.18\% & $\epsilon=\{2.04,8.03\};\delta=10^{-5}$&\{98\%,98.1\%\}&\{1.19\%,1.1\%\}\\
    
    Papernot \emph{et al.} 2017 \cite{Pape17} (\textit{657}) & SVHN & 600,000 & 92.8\% & $\epsilon=\{5.04,8.19\};\delta=10^{-6}$&\{82.7\%,90.7\%\}&\{10.9\%,2.3\%\}\\
    
    Hynes \emph{et al.} 2018 \cite{hynes18} (\textit{68}) & CIFAR10 & 50,000 & 92.4\% & $\epsilon=4;\delta=10^{-5}$ & 90.8\% & 1.72\%\\
    
    Rahman \emph{et al.} 2018 \cite{Rahman18} (\textit{142}) & CIFAR10 & 50,000 & 73.7\% & $\epsilon=\{1,2,4,8\};\delta=\delta=10^{-5}$ & \{25.4\%,45\%,60.7\%,68.1\%\} & \{65.5\%,38.9\%,17.6\%,7.6\%\}\\
    
    Rahman \emph{et al.} 2018 \cite{Rahman18} (\textit{142}) & MNIST & 60,000 & 97\% & $\epsilon=\{1,2,4,8\};\delta=\delta=10^{-5}$ & \{75.7\%,87\%,90.6\%,93.2\%\} & \{21.9\%,10.3\%,6.6\%,3.9\%\}\\
    
    Papernot \emph{et al.} 2021 \cite{Pape21} (\textit{53}) & MNIST & 60,000 & 99\% & $\epsilon=2.93;\delta=10^{-5}$ & 98.1\% & 0.9\%\\
    
    Papernot \emph{et al.} 2021 \cite{Pape21} (\textit{53}) & CIFAR10 & 50,000 & 76.6\% & $\epsilon=7.53;\delta=10^{-5}$ & 66.2\% & 13.6\%\\
    
Huang {\em et al.} 2019~\cite{huang19} (\textit{82}) & Adult & 48,842 & 82\% & $\epsilon=\{0.1,0.5,1.01,2.1\};\delta=10^{-3}$ & $\{55\%,75\%,76\%,77\%\}$ & $\{32.9\%,8.5\%,7.3\%,6.0\%\}$\\
  \bottomrule
\end{tabular}
\end{adjustwidth}
\end{table}

In Table~\ref{tab:ML} we survey the performance of several highly cited works following this approach. Note that the variation in the accuracy obtained for the same data set is caused by the different learning models, number of epochs and train/test sets employed by the various works. Thanks to the use of the moments accountant method, $\epsilon$ values are 
single-digit, although they often reach 8 or above. However, contrary to the widespread opinion in the ML community, 
$\epsilon \in [1..10)$ cannot truly be regarded as safe. For the DP condition (\ref{dpcondition}) to really express small influence of any single record on the DP output, the value of $\epsilon$ must be
small, ``say, $0.01$, $0.1$, or in
some cases, $\ln 2$ or $\ln 3$''~\cite{dwork11}. In Section~\ref{intro}, we showed that $\epsilon=8$ in strict DP amounts to no confidentiality at all in the binary randomized response equivalence. On top of that, all works use values of $\delta$ which are very close or even larger than $1/n$. This implies that strict DP is {\em not} satisfied with a high probability. For such weak privacy parameters, the theoretical guarantees of DP are so diminished that, as stated by the DP inventor~\cite{dwor19}, one can hardly call the outputs DP-protected at all. 

This lack of protection is formally demonstrated in~\cite{yeom2018privacy}, where the authors define the theoretical upper bound on possible leakage in ML for DP guarantees against membership inference attacks (MIAs). In such attacks, an adversary aims to identify the presence or absence of a record in the data set, which is exactly what DP aims to obfuscate. 
Thus, MIAs are the natural way to evaluate the practical guarantees offered by DP in ML~\cite{Jayaraman2019}. 
In~\cite{yeom2018privacy} the authors define the adversary's advantage as the difference between the adversary's true and false positive rates in MIAs, and prove that, if the model satisfies $\epsilon$-DP,
then the adversary's advantage is upper-bounded by $e^\epsilon-1$. For values of $\epsilon>1$, the adversary advantage may be well above 1, which means that the theoretical guarantee does not provide 
real protection. 
As our experiments will show, avoiding ML model overfitting turns out to be more effective against MIAs.

If the theoretical guarantees of DP do not hold, an empirical confidentiality assessment should be carried out ({\em e.g.} by mounting inference attacks). In this respect, \cite{Jayaraman2019} shows that differential privacy relaxations with weak privacy parameters have a practical leakage commensurate {\em with the amount of noise they add} for a given $\epsilon$, rather than with the $\epsilon$ value itself.
This also means that, in practice, protecting ML via DP relaxations boils down to {\em plain} noise addition, a 40-year old technique criticized for its poor privacy/utility
trade-off \cite{Paass1988,Tendick1991}. 

Unfortunately, most works using DP relaxations in ML fail to offer empirical privacy assessments, because any formulation and parameterization connected to DP is often viewed as private enough.

On the other hand, thanks to very relaxed (and unsafe) implementations of DP, the works surveyed in Table~\ref{tab:ML} often achieve moderate accuracy losses. Yet, for the most difficult tasks, losses are 20\% or higher, which is dubiously acceptable if one wants ML to stay useful.

\section{Applying DP in decentralized ML}
\label{decentralized}

Decentralized machine learning (DML) allows training an ML model by a community of peers holding private local data sets. The most popular architecture of DML is federated learning (FL)~\cite{FL}, in which a single server (called model manager) orchestrates the learning process. The model manager sends an initial model to all peers (called clients), who compute model updates on their own private data and return the updates to the model manager. The manager aggregates the updates and re-distributes the new model. Communication iterations carry on until the model converges. Other DML forms exist, such as fully decentralized machine learning (FDML)~\cite{Tang2018}, in which all peers may be managers of their own models and clients for other's models.

DML (and FL specifically) favors privacy because the clients do not need to upload their own private data to the manager. However, the updates returned by the clients to the manager may still leak their private data~\cite{survey}. Moreover, as it happens in centralized ML, the learned model may also leak private data on the participating clients. To prevent such leakages, DP has been employed --again-- as the gold standard to protect clients' data. According to how DP is enforced in FL, 
three types of approaches and subsequent privacy guarantees can be distinguished.

In the first approach the noise addition required for DP is performed locally by each client, either by applying perturbations to the updates resulting from training the local model, or by using DP stochastic gradient descent (DP-SGD) during training~\cite{Abadi2016}.
This approach, known as \emph{local differential privacy} (LDP)~\cite{Naseri2022}, protects the clients' single data points (records) by making their contribution to the learned model unnoticeable. Some works using this approach are~\cite{Pihur2018} and~\cite{Truex2020}; big companies such as Google, Apple and Microsoft also employ this method to systematically collect data of their users~\cite{Ding2018,Erlingsson2014,Thakurta2017}.
Nevertheless, since quite often all the data points of a client refer to the client herself ({\em e.g.} fitness or health measurements over time), they may be highly correlated; hence, LDP may not offer client-level privacy, but just 
\emph{instance-level privacy}~\cite{Triastcyn2019}. In scenarios with clients holding correlated data points, this approach can barely be called DP at all: 
making the presence or absence of a certain data point (instance) unnoticeable does not offer any privacy to the client if that data point is highly correlated to the rest of data points in the client's data set. 

The second approach is more in line with the DP guarantee, and its goal is to hide the presence of any single client in the training process~\cite{Geyer2017}. This method is known as \emph{central differential privacy} (CDP)~\cite{Naseri2022} and 
provides \emph{client-level privacy} guarantees. With CDP, the FL aggregation function is perturbed by the model manager to 
bound the influence of any single client's data on the model parameters. This means that clients need to entrust the manager with their exact model updates, and rely on the manager to correctly
perform the perturbation. Assuming trust in the model manager goes against the purpose of privacy protection, even though it is still better than sharing entire data sets in clear form. Some works using this approach are~\cite{Geyer2017} and~\cite{McMahan2018}.

A third approach is described in~\cite{zhu20}. Each client trains her local model and never reveals it to the manager. When asked to make a prediction on an instance, every client computes a label with her respective local model; then the client adds noise to her label; finally, clients engage in secure multiparty computation to perform a voting among the noise-added labels, so that the majority label is taken as the final label for the 
instance. This approach achieves instance-level R\'enyi DP thanks to the noise added to local labels, and also client-level R\'enyi DP thanks to the voting.

As in centralized ML, DP relaxations are used systematically in FL, with $(\epsilon,\delta)$-DP being the most usual choice. The $\sigma$ parameter is calibrated to bound the privacy loss in each communication round (which requires setting the number of rounds beforehand), and the privacy loss ($\epsilon$) is accumulated via the moments accountant method~\cite{Abadi2016}. The $\delta$ parameter is set according to the number of 
clients participating in the learning process, so that $\delta$ stays well below $1/|clients|$. 

In Table~\ref{tab:FL} we survey the empirical results reported by works employing DP in FL. Deep learning models are the most usual in FL. Hence, we focus on deep learning methods offering 
$(\epsilon,\delta)$-DP and \emph{client-level privacy}, which, as argued above, is the most interesting 
and challenging type of privacy. To simulate a FL scenario, these works split the records in the evaluation data set across a number of simulated
clients. The records held by a client are assumed to describe such client and, therefore, are considered
non-independent data. According to how records are distributed across clients, we can differentiate between independent and identically distributed (i.i.d.) data and non-i.i.d. data, with the latter being the most realistic in FL settings~\cite{Zhu21}. Since results are highly dependent on the number of clients, we report figures on the same data set for various numbers of clients, when available. 

\begin{table}
\tiny
  \caption{DP for federated deep learning: a comparison of recent works}
  \label{tab:FL}
\begin{adjustwidth}{-1.7cm}{}
  \begin{tabular}{llllrrr}
    \toprule
    Reference (\textit{cites})&Data set&|Clients|&Original accuracy&DP parameters&DP accuracy&Relative accuracy loss \\
    \midrule

         Geyer \emph{et al.} 2018 \cite{Geyer2017} (\textit{668}) and & & & & & &\\
     Triastcyn \& Faltings 2019 \cite{Triastcyn2019} (\textit{71})&MNIST (non-i.d.d.)&100&97\%&$\epsilon=8;\delta=10^{-3}$&78\%&19.6\%\\
     
     Geyer \emph{et al.} 2018 \cite{Geyer2017} (\textit{668}) and & & & & & &\\
     Triastcyn \& Faltings 2019 \cite{Triastcyn2019} (\textit{71})&MNIST (non-i.d.d.)&10,000&99\%&$\epsilon=8;\delta=10^{-6}$&96\%&3\%\\
     
    Triastcyn \& Faltings 2019 \cite{Triastcyn2019} (\textit{71})&MNIST (i.i.d.)&100&97\%&$\epsilon=8;\delta=10^{-3}$&86\%&11.3\%\\ 
    
     Triastcyn \& Faltings 2019 \cite{Triastcyn2019} (\textit{71})&MNIST (i.i.d.)&10,000&99\%&$\epsilon=8;\delta=10^{-6}$&97\%&2\%\\ 
     
     Triastcyn \& Faltings 2019 \cite{Triastcyn2019} (\textit{71})&APTOS 2019&100&70\%&$\epsilon=8;\delta=10^{-3}$&60\%&14.3\%\\ 
     
     Triastcyn \& Faltings 2019 \cite{Triastcyn2019} (\textit{71})&APTOS 2019&10,000&72\%&$\epsilon=8;\delta=10^{-6}$&68\%&5.5\%\\ 

    Naseri \emph{et al.} 2022 \cite{Naseri2022} (\textit{41}) &MNIST&100&98\%&$\epsilon=3;\delta=10^{-5}$&82\%&16.3\%\\
    
    Naseri \emph{et al.} 2022 \cite{Naseri2022} (\textit{41}) &CIFAR10&100&93\%&$\epsilon=3;\delta=10^{-5}$&79\%&15\%\\
    
    \bottomrule
\end{tabular}
\end{adjustwidth}
\end{table}

By examining the DP parameters and accuracy values listed in Table \ref{tab:FL} we can extract the following conclusions:
\begin{itemize}
    \item The values of the privacy parameter $\epsilon$ are on the unsafe side, like in Table~\ref{tab:ML} on centralized ML. Therefore, the arguments presented in Section~\ref{centralized} for DP-protected centralized ML also apply to DP-FL.
    \item The number of participating clients has a major impact on the model accuracy for a given privacy level. In general, if the number of clients stays below 1,000, the impact on the model accuracy is significant, up to the point that the model hardly converges \cite{Geyer2017}. Even though DP-FL almost matches the accuracy of non-private FL when the number of clients is around 10,000, this configuration is only realistic for scenarios involving mobile services deployed by big companies such as Google~\cite{Hard2018} or Apple~\cite{Ramaswamy2019}. However, FL is employed in many occasions in much smaller configurations, for example in medical applications~\cite{Brisimi2018};
    unfortunately, resorting to DP-FL to address the privacy requirements of such configurations can lead to unacceptable accuracy loss.
    \item The fact that in FL the
    data might 
    be non-i.i.d. among clients is particularly challenging for privacy, because the different distributions of the local data of the various clients may render some clients easily distinguishable. Consequently, it takes more effort (noise) to hide their contribution to the model, as shown in~\cite{Triastcyn2019}. 
\end{itemize}

In summary, DP-FL only seems to offer competitive accuracy (but still weak privacy guarantees), when the number of clients is exceedingly high and data are i.i.d. among clients. On the one side, these assumptions are often unrealistic. 
On the other side, if there is a great number of clients and their private data are similarly distributed, clients already enjoy some natural privacy protection because they are ``hidden in the crowd'', in which case resorting to non-private FL might be safe enough and certainly simpler.  

\section{Empirical results}
\label{empirical}

As argued in the previous sections, the theoretical privacy guarantees of DP should not be taken for granted in the current implementations of DP-based privacy-preserving ML. In line with what is done in the SDC literature, privacy and utility should be evaluated {\em ex post} by empirically measuring the resistance against membership inference attacks and the accuracy of the models. In this section we report the results of such an empirical evaluation on three data sets commonly employed in the data privacy and ML literature (Adult, MNIST, and CIFAR10) in four different ML settings:
i) a multi-layer perceptron for Adult,
ii) a shallow convolutional model for MNIST,
iii) a deep convolutional model for CIFAR10, and 
iv) a second deep model using transfer learning for CIFAR10.

The experiments were run on an AMD Ryzen 7 5800X
 with 16 GB of RAM
and an NVIDIA GeForce RTX 3070 GPU with 8 GB of VRAM,
using Windows 10,
Python 3.8.10,
and CUDA 11.2.
We used TensorFlow 2.5.0 to build the deep learning models
and the TensorFlow-Privacy 0.6.2 implementations of the
moments accountant DP training algorithm 
and the membership inference attacks.

Next, we describe the model architectures used in each of
the experimental settings. 
The model architecture for the Adult data set consisted of
two fully connected layers of $40$ units each, with 
ReLU as activation function, and an output layer with
two neurons.
For the MNIST data set, we used a model consisting of a 
convolution layer with $12$ channels, kernel size $3$,
and ReLU activation, followed by a max pooling layer and
the output layer, with 10 neurons.
For the CIFAR10 data set, the model architecture included
$2$ sets of convolution and max pooling layers, with $20$ and $50$ channels
respectively, kernel size $5$, and ReLU activation, followed by a fully
connected layer of $500$ neurons with ReLU activation and the output
layer with $10$ neurons.
For the CIFAR10 data set with transfer learning (CIFAR10-TL), the architecture
was composed of the bottom layers of 
DenseNet-201, trained on the ImageNet data set, with frozen weights,
followed by a fully connected layer of $256$ neurons with ReLu activation,
and the output layer with $10$ neurons.
All settings used the categorical cross-entropy loss function and the
Adam optimizer, with learning rate $1.5 \times 10^{-4}$, batch
size of $100$, and $50$ training epochs.

To evaluate privacy, we used the standard black-box membership inference attack implementation
in TensorFlow-Privacy, which is based on the works by Shokri {\em et al.}~\cite{shokri2017membership},
and Salem {\em et al.}~\cite{salem2018ml}. We ran all implemented attack specifications,
using slices including data from all classes,
data from each single class, correctly classified data points, and incorrectly
classified data points.
As proposed in~\cite{yeom2018privacy}, the attack performance (\emph{i.e.}, the inverse of privacy protection) was reported as the maximum attacker advantage (\emph{i.e.}, the difference between the adversary's true and false positive rates 
in MIAs) obtained throughout all attack specifications and slices.
Note, however, that attack-based privacy figures are limited to showing a lower
bound on the information leakage since they measure
the effectiveness of a particular attack. This contrasts with
DP guarantees, which provide an upper bound
on the possible leakage~\cite{Jayaraman2019}.
As discussed in Section~\ref{centralized}, the upper bound for the adversary's advantage is $e^\epsilon-1$, which means that for $\epsilon>1$ the adversary's advantage 
may be well above 1.

For the sake of reproducibility, the code 
of our experiments is available at \url{https://github.com/ablancoj/dp-mia}.

\paragraph{Baseline results} Table~\ref{tab:baseline} reports baseline results for the four experimental
settings with {\em no protection.} We see that, for the three settings other than Adult, the attacker's advantage was around or over 0.5. Also, we see that that in these three cases
there was a big gap between the training and validation losses. Such results are in
line with previous studies~\cite{yeom2018privacy}, which
directly connect the generalization gap (\emph{i.e.}, the difference between training and testing
losses or accuracies) to the vulnerability against both membership and attribute
inference attacks.

\begin{table}
\scriptsize
\centering
\renewcommand{\arraystretch}{1.2}
\caption{Baseline results (no protection)}\label{tab:baseline}
\begin{tabular}{lcccccc}
\toprule
Data set & Training loss & Testing loss & Training accuracy & Testing accuracy & Training time (s) & Attacker's advantage \\
\hline
Adult      & 0.2797 &  0.3685 &  0.8696 &  0.8444 &  46 & 0.131 \\
MNIST      & 0.0066 &  0.0782 &  0.9986 &  0.9808 &  60 & 0.577 \\
CIFAR10    & 0.0505 &  3.4429 &  0.9836 &  0.6743 &  88 & 0.642 \\
CIFAR10-TL & 0.0015 &  0.8154 &  1.0 &     0.8517 &  43 & 0.467 \\
\bottomrule
\end{tabular}
\end{table}

\paragraph{Results using anti-overfitting} To test the relationship between the (excessive) fitting of ML models to the (private) training data and their
vulnerability against membership attacks, we next tested the performance of models when trained using anti-overfitting methods,
namely {\em dropout} and $L_2$-{\em regularization}.
Dropout layers in neural network architectures randomly suppress
connections between neurons at each training epoch according to some 
parameter. On the other hand, $L_2$-regularization adds a term
to the loss function that penalizes the model when it overfits the training
data. These two techniques are used to obtain models that generalize
better to unseen data, and their use is pervasive in machine learning.

Table~\ref{tab:regularization} shows the results for different combinations
of dropout rates and $L_2$-regularization in the four experimental
settings. 
Some highlights of the comparison between 
Tables~\ref{tab:baseline} and~\ref{tab:regularization} follow.
For the Adult data set, training the model with $75\%$ dropout rate and no
$L_2$-regularization reduced the attacker's advantage by approximately
$35\%$ while still improving the test accuracy 
with respect to the baseline. The same parameters reduced
the attacker's advantage by $67\%$ for the MNIST data set, while again improving the test accuracy.
In the CIFAR10 case, $25$\% dropout rate and $L_2$-regularization
improved the test accuracy by more than $4\%$ and reduced the attacker's 
advantage by $84\%$, from $0.642$ to $0.099$.
Finally, when using transfer learning to classify CIFAR10, $25$\% dropout rate
and $L_2$-regularization reduced the attacker's advantage by $71\%$,
while incurring a test accuracy reduction of less than $1\%$.

\begin{table}
\scriptsize
\centering
\renewcommand{\arraystretch}{1.2}
\caption{Results for the four experimental settings with different anti-overfitting techniques}
\label{tab:regularization}
\begin{tabular}{lcccccc}
\toprule
Data set & Dropout & $L_2$ regularization & Training accuracy & Test accuracy & Training time (s) & Attaker's advantage \\
\hline
\multirow{8}{*}{Adult}			& 0.25 & \xmark & 0.8627 &  0.8457 &  53  & 0.132 \\ 
							& 0.50 & \xmark & 0.8522 &  0.8492 &  51  & 0.116 \\
							& 0.75 & \xmark & 0.8353 &  0.8453 &  51  & 0.085 \\
							\cline{2-7} 
							& 0.25 & \cmark & 0.8246 &  0.8235 &  59  & 0.063 \\
							& 0.50 & \cmark & 0.8225 &  0.8255 &  59  & 0.071 \\
							& 0.75 & \cmark & 0.7897 &  0.8137 &  60  & 0.069 \\
							\hline												  
\multirow{8}{*}{MNIST}		& 0.25 & \xmark & 0.9885 &  0.9847 &  61  & 0.454 \\ 
							& 0.50 & \xmark & 0.9768 &  0.9837 &  46  & 0.188 \\
							& 0.75 & \xmark & 0.9503 &  0.9805 &  36  & 0.214 \\
							\cline{2-7} 
							& 0.25 & \cmark & 0.9659 &  0.9725 &  48  & 0.196 \\
							& 0.50 & \cmark & 0.9552 &  0.9758 &  52  & 0.189 \\
							& 0.75 & \cmark & 0.9314 &  0.9714 &  67  & 0.183 \\
							\hline										  
\multirow{8}{*}{CIFAR10}	& 0.25 & \xmark & 0.8762 &  0.7336 &  92  & 0.596 \\ 
							& 0.50 & \xmark & 0.7000 &  0.7368 &  94  & 0.227 \\
							& 0.75 & \xmark & 0.4840 &  0.4796 &  92  & 0.132 \\
							\cline{2-7} 
							& 0.25 & \cmark & 0.6869 &  0.7178 &  97  & 0.099 \\
							& 0.50 & \cmark & 0.6048 &  0.6730 &  101 & 0.136 \\
							& 0.75 & \cmark & 0.4569 &  0.4756 &  101 & 0.111 \\
							\hline						  
\multirow{8}{*}{CIFAR10-TL}	& 0.25 & \xmark & 0.8770 &  0.8611 &  51  & 0.275 \\ 
							& 0.50 & \xmark & 0.7729 &  0.8414 &  50  & 0.120 \\
							& 0.75 & \xmark & 0.6038 &  0.7785 &  49  & 0.135 \\
							\cline{2-7} 
							& 0.25 & \cmark & 0.8177 &  0.8427 &  59  & 0.132 \\
							& 0.50 & \cmark & 0.7542 &  0.8206 &  58  & 0.159 \\
							& 0.75 & \cmark & 0.6300 &  0.7882 &  59  & 0.125 \\
\bottomrule
\end{tabular}
\end{table}

These results confirm that reducing model overfitting  successfully increases
the protection against membership inference attacks while improving in most cases the generalization capabilities of trained models (which is actually the goal of anti-overfitting methods).
Note also that applying dropout and $L_2$-regularization did not have a meaningful impact on the training times.
We also see that employing transfer learning (plus some mild anti-overfitting) is an effective method to
obtain good accuracy in difficult tasks while keeping disclosure at bay, because the model will be less fitted to the (private) training data.

\paragraph{Results using DP} Next, we empirically evaluated the protection against membership inference
attacks afforded by DP when applied to the baseline models.
Table~\ref{tab:dpresults} shows the results for the four experimental
settings when trained under $(\epsilon,\delta)$-DP, in particular DP-SGD
using the moments accountant, as done in all the surveyed works. The value for delta was fixed at $\delta=10^{-6}$,
which is one order of magnitude smaller than that used in related works and more in line with its original formulation,
that is, $\delta \ll 1/n$, where $n$ is the number of records.
The value of $\epsilon$ was set to reproduce the settings commonly employed in the ML literature (in the range $[2,8]$), 
as well as safer configurations (in the range $[0.1,1]$) and weaker configurations (in the range $[8,1000]$). 
The batch size for the Adult data set was set to $48$, and to $100$ for the rest
of the experimental settings (note that TensorFlow-Privacy requires the batch size
to divide the data set size). The gradients were clipped at a maximum norm $2.5$ in all
cases, and the noise multiplier $\sigma$ was computed using the TensorFlow-Privacy
\texttt{compute\_noise} function for several target $\epsilon$ values.
Table~\ref{tab:dpresults} also shows the noise multiplier $\sigma$ and the effective $\epsilon$ value
computed by the moments accountant method after the training process was finished. As we discuss below, getting the desired $\epsilon$ is not trivial and, for extreme values, it may not even be possible.

At a first glance, it is clear that DP significantly reduced
the attacker's advantage compared to the baseline results for all $\epsilon$ values.
The reduction was in line with the results obtained with anti-overfitting techniques.
However, the (test) accuracy was also highly impacted: even for the weakest values of $\epsilon$, test
accuracies were lower than those obtained with a reasonable amount of anti-overfitting.
This suggests that, even though the smallest distortion added by DP was enough to counter
practical membership attacks, such a distortion was more harmful for the learning process than
the meaningful changes introduced by anti-overfitting techniques. In fact, since anti-overfitting changes were specifically tailored to enable the model to generalize better, they also produced the same effect
as DP-ML, namely to make the model more agnostic about the (private) training data.
On the other hand, the accuracy losses for safe values of $\epsilon$ (1 or less) were unacceptable, which seems to indicate that
``true'' DP is unsuitable to achieve privacy protection while preserving usable accuracy in ML. 
To make things worse, note that the training times for DP are much higher than
those of the baseline and the anti-overfitting scenarios.

\begin{table}
\scriptsize
\centering
\renewcommand{\arraystretch}{1.2}
\caption{Results for the four experimental settings using differential privacy}
\label{tab:dpresults}
\begin{tabular}{lccccccc}
\toprule
Data set & Target $\epsilon$ & $\sigma$ & Computed $\epsilon$ & Training accuracy & Test accuracy & Training time (s) & Attacker's advantage \\ 
\hline
\multirow{9}{*}{Adult}      & 0.1  & 14.3098 & 0.1120 & 0.7498 & 0.7535 & 774  & 0.083 \\
                            & 0.5  & 3.0864  & 0.5000 & 0.8076 & 0.8016 & 784  & 0.078 \\
                            & 1    & 1.7217  & 0.9999 & 0.8186 & 0.8090 & 741  & 0.082 \\
                            & 2    & 1.0674  & 2.0000 & 0.8225 & 0.8190 & 781  & 0.077 \\
                            & 4    & 0.7767  & 3.9969 & 0.8201 & 0.8132 & 790  & 0.062 \\
                            & 8    & 0.6223  & 7.9999 & 0.8228 & 0.8201 & 850  & 0.099 \\
                            & 16   & 0.5172  & 15.922 & 0.8250 & 0.8201 & 778  & 0.061 \\
                            & 100  & 0.3320  & 97.440 & 0.8294 & 0.8234 & 754  & 0.071 \\
                            & 1,000 & 0.2183  & 484.87 & 0.8318 & 0.8283 & 854  & 0.079 \\
\hline																					  
\multirow{9}{*}{MNIST}      & 0.1  & 11.9929 & 0.1120 & 0.8388 & 0.8514 & 2,555 & 0.201 \\
                            & 0.5  & 2.6130  & 0.4999 & 0.8912 & 0.9009 & 2,308 & 0.17  \\
                            & 1    & 1.4877  & 0.9999 & 0.8990 & 0.9086 & 2,460 & 0.153 \\
                            & 2    & 0.9675  & 1.9851 & 0.9052 & 0.9110 & 2,457 & 0.165 \\
                            & 4    & 0.7327  & 3.9161 & 0.9085 & 0.9154 & 2,311 & 0.169 \\
                            & 8    & 0.5922  & 7.9737 & 0.9114 & 0.9172 & 2,459 & 0.19  \\
                            & 16   & 0.4958  & 15.953 & 0.9154 & 0.9208 & 2,312 & 0.179 \\
                            & 100  & 0.3234  & 97.604 & 0.9172 & 0.9217 & 2,311 & 0.191 \\
                            & 1,000 & 0.2155  & 482.07 & 0.9294 & 0.9337 & 2,312 & 0.188 \\				 
\hline																					  
\multirow{9}{*}{CIFAR10}    & 0.1  & 13.1337 & 0.1120 & 0.2624 & 0.2662 & 3,796 & 0.12  \\
                            & 0.5  & 2.8456  & 0.4999 & 0.3166 & 0.3160 & 3,832 & 0.12  \\
                            & 1    & 1.6022  & 0.9999 & 0.3753 & 0.3777 & 3,830 & 0.10  \\
                            & 2    & 1.0134  & 2.0000 & 0.4110 & 0.4068 & 3,821 & 0.13  \\
                            & 4    & 0.7533  & 3.9718 & 0.4177 & 0.4181 & 3,791 & 0.15  \\
                            & 8    & 0.6070  & 7.9956 & 0.4399 & 0.4397 & 3,819 & 0.09  \\
                            & 16   & 0.5065  & 15.936 & 0.4502 & 0.4435 & 3,918 & 0.12  \\
                            & 100  & 0.3277  & 97.520 & 0.4895 & 0.4857 & 4,093 & 0.12  \\
                            & 1,000 & 0.2169  & 483.49 & 0.5109 & 0.4988 & 4,255 & 0.1   \\
\hline																					  
\multirow{9}{*}{CIFAR10-TL} & 0.1  & 13.1337 & 0.1120 & 0.6535 & 0.6535 & 1,605 & 0.112 \\
                            & 0.5  & 2.8456  & 0.4999 & 0.7723 & 0.7577 & 1,620 & 0.124 \\
                            & 1    & 1.6022  & 0.9999 & 0.7979 & 0.7911 & 1,627 & 0.16  \\
                            & 2    & 1.0134  & 2.0000 & 0.8149 & 0.8030 & 1,618 & 0.123 \\
                            & 4    & 0.7533  & 3.9718 & 0.8247 & 0.8086 & 1,615 & 0.14  \\
                            & 8    & 0.6070  & 7.9956 & 0.8327 & 0.8169 & 1,728 & 0.128 \\
                            & 16   & 0.5065  & 15.936 & 0.8385 & 0.8180 & 1,652 & 0.143 \\
                            & 100  & 0.3277  & 97.520 & 0.8493 & 0.8280 & 1,612 & 0.136 \\
                            & 1,000 & 0.2169  & 483.49 & 0.8615 & 0.8377 & 1,614 & 0.128 \\
\bottomrule
\end{tabular}
\end{table}

In fact, a suboptimal privacy/utility trade-off is not the only issue faced when using DP-SGD.
The choice of hyperparameters to train a model under DP is a challenging
task, as we discuss in the what follows. 
The \texttt{compute\_noise} function in TensorFlow-Privacy allows computing the noise 
multiplier $\sigma$ required to achieve a certain target $\epsilon$, given a fixed $\delta$, the size
of the training data set, the batch size, and the expected number of training epochs. 
This noise multiplier is then used to compute the noise added to each of the model gradients
at each training epoch. Then, the \texttt{get\_privacy\_spent} function returns the 
effective $\epsilon$ value computed by the moments accountant after the training is done.
However, as introduced in Section~\ref{fitting}, if a model is trained until convergence (as is usually the case),
the computation of the noise multiplier is unreliable ---if the model is finally trained
for more epochs than expected, the final effective $\epsilon$ will be greater than intended,
which means that more noise should have been added at each training epoch to achieve the
target $\epsilon$. But, in turn, adding more noise might require more training epochs for the model to converge. 
On the other hand, if the model converges before the number of epochs specified when computing $\sigma$,
the effective $\epsilon$ will be smaller than intended, which means too much noise was added during training, with the subsequent potential (and unnecessary) loss of model accuracy.  

The above-described pitfalls of hyperparameter choice are exemplified in Table~\ref{tab:dpconverge}. In this experiment we trained a
model to classify CIFAR10 until convergence, that is, until the training loss did not
significantly change for $3$ consecutive epochs. 
The target privacy budget $\epsilon$ was set to $0.1$.
We initially computed the required noise multiplier $\sigma$ by using \texttt{compute\_noise}
and specifying $10$ training epochs.
The model turned out to converge in $72$ epochs, with a computed effective $\epsilon=0.2704$.
Next, we re-computed $\sigma$ by specifying $72$ training epochs and
we trained again until convergence.
However, since the amount of noise added per epoch had changed, so did the epochs needed to attain convergence: just 12 epochs were needed this time. 
This required repeating the process until an effective $\epsilon=0.0999$ was obtained, which was deemed close enough to the target $\epsilon=0.1$.
This trial and error process is similar to the utility-first approach traditionally followed by official statisticians since the 1970s, well before privacy models and DP were introduced:
choose noise parameters, protect the data using an SDC method with those parameters, and check if the protected data
are accurate and private enough according to some requirements. 

\begin{table}
\scriptsize
\centering
\renewcommand{\arraystretch}{1.2}
\caption{Iterative process to obtain the desired $\epsilon$ value of $0.1$}
\label{tab:dpconverge}
\begin{tabular}{ccccc}
\toprule
Target training epochs & $\sigma$ & Epochs to converge & $\epsilon$ & Test accuracy \\
\midrule
10 &  5.92&  72 & 0.2704 & 0.25 \\
72 &  15.75& 12 & 0.0878 & 0.1  \\
12 &  6.478& 85 & 0.2681 & 0.26 \\
85 &  17.10& 40 & 0.0966 & 0.15 \\
40 &  11.75& 48 & 0.1179 & 0.18 \\
48 &  12.87& 28 & 0.0999 & 0.12 \\
\bottomrule
\end{tabular}
\end{table}

Furthermore, DP-SGD does not scale well for complex models, as it causes their accuracy to cap at approximately 1 over the number of  classes ({\em i.e.}, random guessing). This can be sometimes solved by
leveraging transfer learning. In that case, the bottom layers (or feature extractor) are
already trained using some public data set, and only the top layers (the classifier) are trained
using DP-SGD.

Finally, the DP-SGD algorithm requires gradients to be clipped before any noise is added to them on a
per-training-instance basis, which eliminates the performance gains of using GPUs to process the training
data in batches. Whereas the \texttt{microbatches} variable in TensorFlow-Privacy can be adjusted to treat the
training instances in bigger batches, which reduces the training time, it does so at the cost of a greater
accuracy loss compared to the baseline results.
Giving up batches explains why the training times of DP-SGD reported in Table~\ref{tab:dpresults} are 15 to 40 times longer 
than those of the baseline and anti-overfitting experiments reported in Tables~\ref{tab:baseline}
and~\ref{tab:regularization}.

Even if the experiments reported in this section are for centralized ML, the conclusions reached can be expected to be valid for FL as well. Note that guaranteeing privacy in FL can be even more challenging, because membership attacks may be easier if the clients' local data are non-i.i.d.

\section{Conclusions}
\label{conclusions}

The ML research community have embraced DP as the gold standard for privacy protection. Yet, none of the related works in the literature employs the original formulation of DP, but relaxations of it, such as $(\epsilon,\delta)$-DP and others. On top of that, the privacy parameters used are quite far from what is considered theoretically ``safe'' ({\em i.e.}, $\epsilon \leq 1$ and negligible $\delta$ w.r.t. the population size). For such loose implementations, the 
privacy guarantees of DP do not hold, and what is done can only be viewed as old plain noise addition, which at best provides {\em ex post} protection depending on the amount of noise added for a given $\epsilon$, rather than on the $\epsilon$ value itself. This renders it rather pointless to claim use of the DP privacy model and comes close to the (often criticized) utility-first paradigm used in SDC for decades. Thus, the use of the ``differential privacy'' label by a lot of works whose implementations offer an actual privacy very far from truly sound DP can only be regarded as deceptive. To aggravate the matter, most of those works systematically omit empirical privacy analyses, because they consider any relaxation and parameterization of DP to be privacy-preserving enough.

In spite of all the aforementioned DP ``twisting and bending'', the practical accuracy/privacy trade-off of DP-ML is worse than that of standard methods used in ML to mitigate overfitting. Our experiments indicate that the latter provide comparable practical protection, better accuracy and significantly lower learning cost than DP-based ML.
On the other hand, even with DP relaxations and adjusted privacy accounting methods, using theoretically sound privacy parameters ($\epsilon \leq 1$) is still infeasible, due to the unacceptably high accuracy loss they entail. This makes ``true'' DP unsuitable for protecting ML models with usable accuracy.

\section*{Acknowledgments}

Partial support to this work has been received from the European Commission
(projects H2020-871042 ``SoBigData++'' and H2020-101006879
``MobiDataLab''), the Norwegian Research Council (project no. 308904
``CLEANUP''),
the Government of Catalonia (ICREA Acad\`emia Prizes to J. Domingo-Ferrer
and D. S\'anchez), the UK Research and Innovation (project MC\_PC\_21033
``GRAIMatter''), and MCIN/AEI /10.13039/501100011033 /FEDER, UE
 (project PID2021-123637NB-I00 ``CURLING''). The opinions in this paper
are the authors' own and do not commit UNESCO or any of the funders.

\bibliographystyle{ACM-Reference-Format}
\bibliography{acmart}

\end{document}